\documentclass{pasj01}
\Received{$\langle$reception date$\rangle$}
\Accepted{$\langle$acception date$\rangle$}
\Published{$\langle$publication date$\rangle$}

\begin{document}

\title{Suzaku observation of a high entropy cluster Abell 548W}
\author{Kazuhiro Nakazawa\altaffilmark{1}, Yuichi Kato\altaffilmark{1}, Liyi Gu\altaffilmark{2}, Madoka Kawaharada\altaffilmark{3}, Motokazu Takizawa\altaffilmark{4}, Yutaka Fujita\altaffilmark{5} and Kazuo Makishima\altaffilmark{1,6}}
\altaffiltext{1}{Department of Physics, the University of Tokyo, 7-3-1 Hongo, Bunkyo-ku, Tokyo 113--0033 }
\altaffiltext{2}{SRON Netherlands Institute for Space Research, Sorbonnelaan 2, 3584 CA Utrecht, The Netherlands}
\altaffiltext{3}{Tsukuba Space Center, Japan Aerospace Exploration Agency, 2-1-1 Sengen, Tsukuba, Ibaraki 305-8505, Japan}
\altaffiltext{4}{Department of Physics, Yamagata University, 1-4-12 Kojirakawa-machi, Yamagata 990-8560}
\altaffiltext{5}{Department of Earth and Space Science, Graduate School of Science, Osaka University, 1-1 Machikaneyama-cho, Toyonaka, Osaka 560-0043, Japan}
\altaffiltext{6}{MAXI Team, Institute of Physical and Chemical Research, 2-1 Hirosawa, Wako, Saitama 351-0198}
\email{nakazawa@juno.phys.s.u-tokyo.ac.jp}

\maketitle

\begin{abstract}
Abell 548W, one of the galaxy clusters located in the Abell 548 region, has about an order of magnitude lower
X-ray luminosity compared to ordinal clusters in view of the well known intracluster medium (ICM)
temperature vs X-ray luminosity ($kT$-$L_X$) relation.
The cluster hosts a pair of diffuse radio sources to the north west and north, both about $10'$ apart from the cluster center.
They are candidate radio relics, frequently associated with merging clusters.
A Suzaku deep observation with exposure of 84.4~ks was performed to search signatures for merging in this cluster.
The XIS detectors successfully detected the ICM emission out to $16'$ from the cluster center.
The temperature is $\sim 3.6$~keV around its center, and $\sim 2$~keV at the outermost regions.
The hot region ($\sim 6$~keV) aside the relic candidates shifted to the cluster center reported by XMM-Newton was not 
seen in the Suzaku data, although its temperature of 3.6~keV itself is higher than the average temperature of 2.5~keV around the radio sources.
In addition, a signature of a cool ($kT \sim 0.9$~keV) component was found around the north west source.
A marginal temperature jump at its outer-edge was also found, consistent with the canonical idea of shock acceleration origin of the radio relics.
The cluster has among the highest central entropy of $\sim 400$~keV cm$^2$ and is 
one of the so-called low surface brightness clusters.
Taking into account the fact that its shape itself is relatively circular and smooth and also its temperature structure is nearly flat,
possible scenarios for merging is discussed.
\end{abstract}

\section{Introduction}

Merging clusters of galaxies is an aspect of gravitational growing of the large scale structure.
While there are many ``apparently circular and relaxed'' clusters,
10--20\% of the clusters show evidences for on-going merger, consistent with 
the dynamical timescale of a cluster ($\sim 10^9$ years) compared to Hubble time ($\sim 10^{10}$ years).
The intracluster medium (ICM), the vast hot plasma emitting X-rays, 
permeating the gravitational potential of the cluster is strongly affected by the merger events.
In X-rays, many mergers have complicated shapes, often strongly elongated, 
with signatures of complex temperature structure. Entropy of the ICM of mergers are often high, 
considered to be due to heating by shock wave as well as mixing of the outer high entropy gas with inner low entropy ones.

Among the 33 flux-limited non-biased samples of REXCESS, 
three (Abell 2399, Abell 3771 and Abell 2328
\footnote{Also named as RXC J2157.4-0747, RXC J2129.8-5048 and RXC J2-48.1-1750, respectively.}) 
show very low surface brightness (LSB), 
indicating high entropy of the ICM. They all lack bright central core, 
apparently elongated and/or have sub-structures, showing 
that they are dynamically young (B\"{o}ringer et al. 2007). 
Another template of this type of cluster is Abell~76, 
which has a very low surface brightness and a complicated structure (Ota et al. 2014).
By the Suzaku observations, its ICM with a temperature of $\sim 3$~keV 
is shown to have very high entropy of $\sim 400$~keV cm$^2$ in its center.
These clusters are outliers in cluster scaling relations, such as 
ICM temperature vs X-ray luminosity ($kT$-$L_X$) relation and their origin is not clear yet. 
Thus, it is important to understand the nature of these LSB clusters as extreme cases.

Abell 548W (or Abell 548b) is one of the 3 major cluster-sized diffuse X-ray sources 
detected in the Abell 548 region (e.g. Davis et al. 1995).
It has a redshift of $z=0.0424$ (Solovyeva et al. 2008), 
with the ICM temperature of $kT \sim 3.6$~keV and
$L_X = 12.6\pm7.0 \times10^{42}~h_{50}^{-2}$ erg s$^{-1}$ cm$^{-2}$ 
(or $6.4\pm3.6 \times10^{42}~h_{70}^{-2}$ erg s$^{-1}$ cm$^{-2}$)  at 0.1--2.4 keV (Davis et al. 1995). 
Here, Hubble constant is $H_0 = 50 \times h_{50}  = 70 \times h_{70}$ km s$^{-1}$ Mpc$^{-1}$.
Notably, the luminosity is an order of magnitude smaller 
compared to other clusters with similar temperature (see, e.g. $kT$-$L_X$ relation figure 12  of Fukazawa et al. 2004).
This property make Abell 548W a typical LSB cluster.

There are two bright elliptical galaxies in its center, $30''$ apart, and the X-ray peaks possibly associated with them.
On the other hand, its X-ray morphology is in general circular, distinct from other LSB clusters.
The cluster is known to have high velocity dispersion of $\sigma_v = 1300$~km s$^{-1}$ (Solovyeva et al. 2008).
From the well known $kT$-$\sigma$ relation $kT = (\frac{\sigma_v}{323.6~{\rm km~s}^{-1}})^{1.49}$ (Xue et al. 2000),
this indicates $kT \sim 8$~keV which is apparently too high.
Solovyeva et al. (2008) attributed this inconsistency as a result of line-of-sight merger
with a velocity shift of $\sim 1500$ km s$^{-1}$.

In the north west (NW) and north directions, $7'$--$13'$ apart from the cluster center, there are two diffuse radio emission
observed at 1.4 GHz (Feretti et al. 2006). A source to the NW
has a flux density of $61\pm5$ mJy at 1.4 GHz and the source to the north $88\pm6$ Jy.
Both sources are polarized by $\sim 30$\% and have a steep spectra of $\alpha = -2\pm1$. 
Their origin is not clear, but these properties as well as apparent non-association with any galaxy 
make them good candidates as cluster radio relics (Feretti et al. 2006).

By analyzing the XMM-Newton (hereafter XMM) data, Solovyeva et al. (2008) reported a
hot region located at $r = 4'$--$7'$ from the cluster center, 
aside (and not within) the relic candidate regions.
In merger scenario, hot region is in many cases coincident in position with relics
(e.g. Akamatsu and Kawahara 2013), with only a few exceptions (e.g. Ogrean et al. 2013), 
which make this result rather confusing.
To explain the results by XMM, the authors made a merger model, 
in which a part of the shocked region shows radio emission while other parts do not, 
and these two regions are slightly overlapping, i.e. a rather complicated geometry.

To resolve the dynamical status of this cluster, a high-sensitivity observation out to the relic region and farther is needed.
In this paper, we revisited this issue with Suzaku (Mitsuda et al. 2007)
utilizing its high sensitivity for low surface brightness diffuse emission.  
We use $H_0 = 70$ km s$^{-1}$ Mpc$^{-1}$ in the following sections.
Distance to the cluster is thus 182~Mpc, and $1'$ angular distance corresponds to 55.2~kpc.
The solar abundances are normalized to those of Anders and Grevesse (1989). 
Observation parameters and obtained image is discussed in section 2, spectral analysis in section 3,
followed by discussion (section 4) and summary (section 5). Otherwise noted, 
all the error-bars are shown in 90\% confidence level.

\section{Observation and data reduction}

Using Suzaku, we observed this source from 14 to 16 February 2013 in a single pointing aimed at
(Ra, Dec) = (86.233$^{\circ}$,$-25.841^{\circ}$), a little offset to NW from the cluster center
to locate the two relics near the center of the field of view (FOV).
The XIS instrument (Koyama et al. 2007) was operated in the full window mode with the spaced-raw charge injection 
(Uchiyama et al. 2009). The data was processed via revision 2.8 pipe-line, and screened with
nominal parameters as follows: not within nor right after the South Atlantic Anomaly (SAA\_HXD $= 0$ and T\_SAA\_HXD $> 436$~s),
apart from dark and sun-lit earth (ELV $> 5^{\circ}$ and DYE\_ELV $>20^{\circ}$), 
and not within low geomagnetic cutoff rigidity region (COR $>6$ GV c$^{-1}$). 
Total effective exposure thus obtained was 84.4~ks.

\begin{figure}[htbp]
 \begin{center}
  \includegraphics[width=8cm]{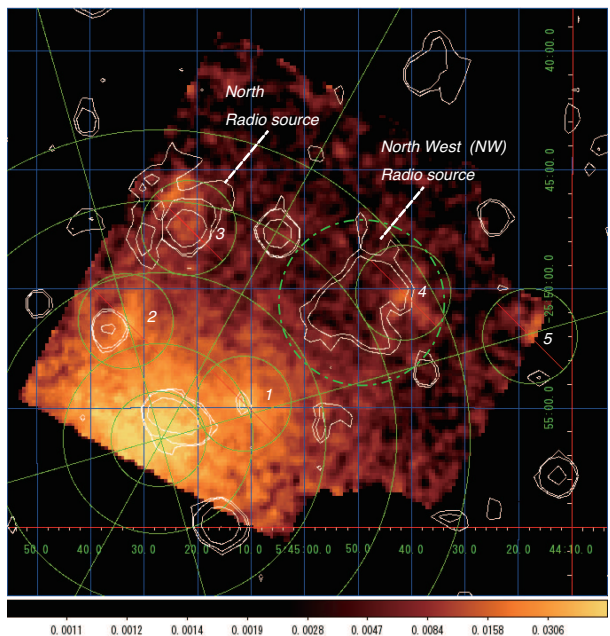}
 \end{center}
 \caption{An X-ray image of Abell 548W in the 0.7--7.0~keV band obtained from XIS0 and 3, after subtracting the NXB and corrected for vignetting and exposure. The 1.4~GHz radio image contours by NVSS survey in white is overlaid to show the location of diffuse radio sources. Green lines defines the spectral analysis regions, with annular radius of $2'$, $4'$, $7'$, $10'$, $13'$ and further out, together with the extraction region of the 5 point sources. Green dot-dashed line represent the region $3'.5$ around the NW relic analyzed in subsection \ref{sec:IC}}\
\label{fig:xis03image}
\end{figure}

The X-ray events of the two front-illuminated (FI) CCDs (XIS0 and XIS3) are combined 
to obtain an image in the 0.7--7.0~keV band, as shown in figure \ref{fig:xis03image}.
Here, the non-X-ray background (NXB) image was produced using the software {\it xisnxbgen} (Tawa et al. 2008).
Vignetting effect is corrected using the flat image made by {\it xissim} (Ishisaki et al. 2007).
The ICM emission was apparently detected out to the relic position. 
Actually, from the following spectral analysis, we detect ICM X-rays out to $\sim 16'$ from the cluster center
(see the next section).

In the image, we identified 5 contaminating point sources clearly visible.
They are also visible in the XMM data, and we discarded regions $r < 2'$ from these sources.
The source nearest to the center $\sim 4'$ offset to the west (source 1) is a hard source presumably an AGN
associated with weak radio signals. Another one $\sim 5'$ to the north (source 2) has similar properties.
Unfortunately, there is a source located very near to the north relic candidate (source 3)
and another one near the edge of the NW one (source 4).
If we include the signal from these sources in the following spectral analysis, both spectra get significantly harder,
so exclusion of the region around these sources are necessary.
The 5th source located $\sim 13'$ to the west is also masked out.
From the XMM data, these sources are shown to have a flux from  
$5.4\times10^{-13}$ (source 1) to $1.3\times10^{-14}$~erg~cm$^{-2}$~s$^{-1}$ (source 5) in the 2--10 keV band.

\section{Spectral analysis}
\label{chap:spectra}

After masking out these 5 sources, 
we defined regions with annular radius separated at $r =2'$, $4'$, $7'$, $10'$,  $13'$ and further out to $\sim 16'$.
Central region is circular, while the other regions are made of two $45^{\circ}$ opening arcs, oriented to the north and NW relic candidates.
The center cordinate follows that of the REFLEX catalog using {\it ROSAT} data (B\"{o}hringer et al. 2004).
The NXB is generated and subtracted using {\it xisnxbgen} again, 
while the CXB is modeled using the flat arf made from {\it xissimarfgen} (Ishisaki et al. 2007).

In our spectral analysis, the new approach to handle the increasing flickering pixel in the NXB template is applied\footnote{see ``A new recipe for generating NXB background spectra. (2015-04-24 by the XIS team)'',  http://www.astro.isas.jaxa.jp/suzaku/analysis/xis/nxb\_new/}. Because the NXB template has a longer exposure, the flickering pixel detection is more sensitive and hence their number is larger than those detected from a single observation. The method applies the flickering pixel lists generated from the NXB database to both the NXB template and the observation data.  

At the time of this paper writing, effect of this ``additional flickering pixel'' is not handled in the effective area estimation by {\it xissimarfgen} and we need to correct this effect manually. It can be approximately estimated by comparing the photon counts before and after applying this new method with reasonable accuracy. It was as small as 0.5--3.5\% for the front illuminated (FI) CCDs (XIS0 and 3), while was as large as 7--20\% for the back-illuminated (BI) CCD (XIS1). Thus, we simply scaled the normalization of the FI CCD data to this difference ratio (i.e. 0.5--3.5\%), while letting the XIS1 normalization to be free.

In Suzaku X-ray spectra, it is known that there are two foreground soft components in addition to the CXB:
the local hot bubble (LHB) modeled by a thermal emission with a temperature $kT = 0.8$~keV,
and another thermal $kT = 0.1\sim0.4$~keV component called the milky way halo (MHW).
To estimate these celestial background component, we first fitted the spectra of the 
two outermost regions ($13'< r < 16'$, north and NW) simultaneously using  
a thermal component with $kT = 0.8$~keV (using {\it apec} code in {\it xspec}), 
another thermal component with free $kT$ and a $\Gamma = 1.41$ fixed power-law.
The third component were modified with a fixed galactic absorption of 
$N_H = 0.0139$~cm$^{-2}$ derived using the {\it w3nh} service from NASA (Dickey \& Lockman 1990).\footnote{NASA W3NH service: https://heasarc.gsfc.nasa.gov/cgi-bin/Tools/w3nh/w3nh.pl }.
Here we used the co-added spectra of XIS0 and 3 (hereafter FI spectra) in the 0.7--8.0~keV band,
as well as those of the BI CCD (XIS1) in the 0.5--6.0~keV band.
The fit became acceptable with $\chi^2/{\rm dof} = 167.7/151$, 
although the residual spectra showed clear trend for softer component. 
When the photon index was left free, it became $\Gamma =$ 1.68--1.85,  which is inconsistent with canonically known CXB spectra.
The derived power-law flux in 2--10 keV band was
$7.18\times 10^{-8}$~erg cm$^{-2}$ s$^{-1}$ str$^{-1}$, 
which is 28.4\% higher than the value obtained in the Suzaku Lockman-hole observation (ID 101002010).

The CXB intensity fluctuates field to field and its level can be approximated by assuming a certain source distribution.
Following the method  used in Nakazawa et al. (2009), which itself is based on Kushino et al. (2002),
we estimated the 90\% confidence fluctuation level for this region. 
Assuming the source cut flux of $S_c \sim 1\times10^{-14}$~erg cm$^{-2}$ s$^{-1}$ at 2--10 keV,
combined with the region area of 0.0158 degree$^2$, we get a number of 15.5\%, which cannot explain the observed difference.
Combined with the soft spectra, we thus conclude that even in these outermost region, the ICM component from Abell~548W is detected.

Because the ICM emission is contaminating out to the outermost region, 
we need to estimate appropriate CXB level by other information.
Analysis of another Suzaku observation 6.6 degree south to Abell 548W (ID 405059010), 
after excluding the central soft source, showed a CXB normalization within 0.6\% to those of the Lockman-hole observation.
We hence utilized this value as a base-line, i.e. $5.58\times10^{-8}$~erg cm$^{-2}$ s$^{-1}$ str$^{-1}$, 
and handled the fluctuation separately.
This gave the parameters of the two foreground components (LHB and MWH) as listed in table~\ref{tab:foreground}. 
For  all the inner regions, we used this value as a fixed foreground component.

\begin{table}
\caption{Fitted results to the foreground components.}
\label{tab:foreground}
\begin{center}
\begin{tabular}{lll}
\hline
components & temperature (keV) & normalization \\
\hline
LHB		& 0.08 (fixed) & $0.84_{-0.84}^{+1.11}$ \\
MWH	& $0.18_{-0.04}^{+0.09}$ & $0.08_{-0.06}^{+0.12}$\\
\hline
\end{tabular}
\end{center}
\footnotesize{normalization in {\it apec} model, scaled to $\pi (20)^2$~arcmin$^2$ flat region.}
\end{table}

\begin{figure}
\begin{center}
  \includegraphics[width=11cm]{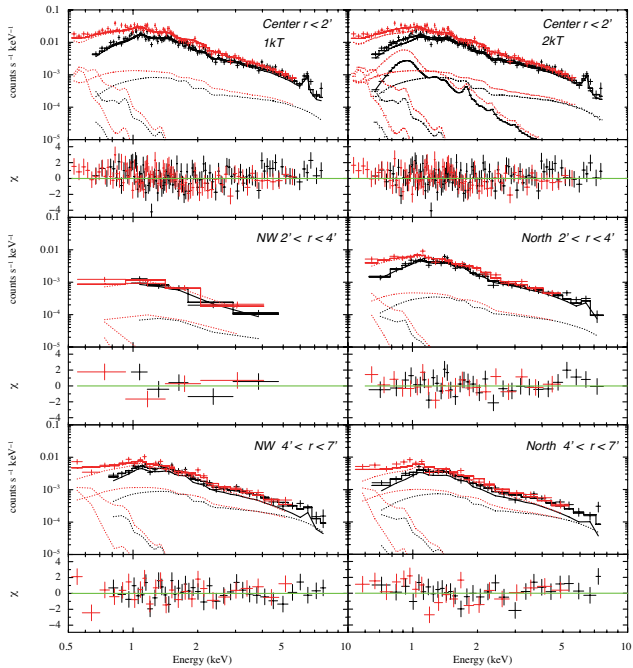}
 \end{center}
 \caption{NXB subtracted XIS spectra fitted with the 1kT model of the inner 5 regions. Black crosses are from the summed XIS0 and 3 (or FI) spectra and red ones are from XIS1. Spectral model consists of the two foreground components (LHB and MWH), the CXB, and the ICM emission. For clarity, the ICM component in the FI models are shown with solid lines, while all the other model components are in dotted lines. Top right panel stands for the 2kT model results for the center spectra. See text for details. }
 \label{fig:spectra1}
\end{figure}

\begin{figure}
 \begin{center}
  \includegraphics[width=11cm]{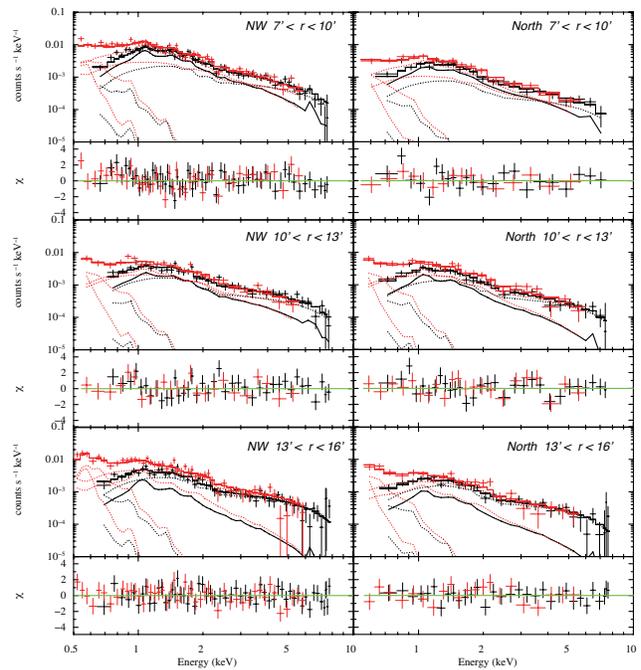}
 \end{center}
 \caption{The same as figure \ref{fig:spectra1}, but for the outer 6 regions. }
 \label{fig:spectra2}
\end{figure}

With the CXB and the two foreground components derived, we fitted all spectra with a single temperature thermal 
emission using the {\it apec} code (hereafter 1kT model).
Systematic uncertainties of the NXB are assumed to be 
2.1\% for the FI spectra and 4.9\% for the XIS1, following table 7 of Tawa et al. (2008), 
which stands for ``5--12 keV NXB reproducibility for 50~ks exposure bins'', modified to 90\% confidence limit. 
Note that effect of NXB reproducibility is minor, as shown in the following results.
Fluctuation of the CXB is calculated in the same way as described above.
All spectra are shown in figure \ref{fig:spectra1} and \ref{fig:spectra2} and fitted results are listed in table \ref{tab:spectra}. In the fitting, 
redshift is fixed at $z = 0.0424$ optically estimated by Solovyeva et al. (2008). Abundance ($A$) is also 
fixed at 0.3 of the solar value, a typical value for a cluster.
In our Suzaku data, spectra of the central $r<2'$ shows clear Fe-K line, while in the other regions it is not clear.
If we set $z$ and $A$ free in the former fitting, they are derived as $z = 0.047^{+0.009}_{-0.006}$ 
and $A = 0.36_{-0.09}^{+0.10}$, respectively, both of which are consistent with the assumed value.

Among the 11 fit results, the central region showed 
relatively large $\chi^2$ resulting in low null-hypothesis probability (NHP) of 0.4\%,
with clear concave residual suggesting its 
multi-temperature nature. When fitted with two temperature thermal emission (hereafter 2kT model), the fit significantly improved with 
the $f$-statistics probability of 0.06\%. Fit results are shown in the top right panel of figure \ref{fig:spectra1} and table \ref{tab:spectra2}.  
The hotter component dominates the spectra and its parameters are not much different from the 1kT fit results 
(e.g. only 0.42 keV higher in temperature and 8\% smaller in normalization),
while the cooler component is minor (only 6\% of the hot component normalization) and will be naturally explained as the inter-stellar medium 
(ISM) of the central galaxies. Hereafter, we interpret the hotter component as the ICM.

While the fits to other 8 spectra are acceptable in 90\% confidence level, the remaining 2 spectra, 
NW $2'<r<4'$ and north $4'<r<7'$, showed marginally poor fit with NHP of 6.3\% and 9.9\%, respectively. 
Since the former one has very low statistics (due to masking out source 1), we could not go into its detail. 
The latter is almost acceptable in 90\% confidence and thus we stick to the 1kT fit for a moment.
Later in subsection 4.2 we will revisit the spectra. 

\small
\def\arraystretch{1.25}
\begin{longtable}{l  lll }
\caption{Temperature and normalization of the spectra from the 11 regions with the 1kT model fit.}
\label{tab:spectra}
\hline
\hline
 & \multicolumn{3}{l}{Center}\\
Region	 & $kT$ keV $^1$ & {\it norm} $^2$& $\chi^2$/dof \\
\hline
$r<2'$ & $3.57 \pm 0.23 \pm 0.10 \pm0.02$  & $6.32\pm 0.19 \pm 0.14 \pm 0.00 \times10^{-2}$ & 223.8/179 \\
\hline
\hline
 & \multicolumn{3}{l}{NW}  \\
Region	 & $kT$ keV $^1$ & {\it norm} $^2$& $\chi^2$/dof \\
\hline
$2'<r<4'$& $3.27_{-0.86}^{+1.79}$ $_{-0.19}^{+0.20} \pm{0.02}$  & $3.77_{-0.61}^{+0.63} \pm 0.23 \pm 0.00 \times10^{-2}$ & 11.9/6  \\
$4'<r<7'$ & $3.60_{-0.41}^{+0.53}$ $_{-0.35}^{+0.37} \pm 0.05$ & $1.63\pm 0.11 \pm 0.16 \pm 0.00 \times10^{-2}$ & 62.7/61  \\
$7'<r<10'$ & $2.52_{-0.26}^{+0.29}$ $_{-0.36}^{+0.35} \pm 0.04$  & $8.38\pm 0.06 \pm 0.10 \pm 0.00 \times10^{-2}$ & 105.0/100 \\
$10'<r<13'$ & $3.32_{-0.71}^{+1.15}$$_{-0.88}^{+0.85}$$_{-0.12}^{+0.20}$ & $0.45 \pm 0.05 \pm 0.11 \pm 0.00 \times10^{-2}$ & 61.4/60 \\
$13'<r<16'$ & $1.81_{-0.28}^{+0.48}$$_{-0.39}^{+0.56}$$_{-0.09}^{+0.10}$ &$0.18 \pm 0.04$$_{-0.08}^{+0.09}$$_{-0.00}^{+0.01} \times10^{-2}$ & 117.5/148$^3$ \\
\hline
\hline
& \multicolumn{3}{l}{North}  \\
Region 	 & $kT$ keV $^1$ & {\it norm} $^2$& $\chi^2$/dof \\
\hline
$2'<r<4'$& $ 4.27_{-0.48}^{+0.78}$$_{-0.19}^{+0.32} \pm 0.02$  & $4.37_{-0.27}^{+0.26}$$_{-0.23}^{+0.22} \pm 0.00 \times10^{-2}$  & 45.4/51\\
$4'<r<7'$ &$3.54_{-0.58}^{+0.66}$$_{-0.50}^{+0.42} \pm 0.05$  & $1.47_{-0.11}^{+0.12}$$_{-0.16}^{+0.17} \pm 0.00 \times10^{-2}$ & 61.0/48 \\
$7'<r<10'$ & $3.01_{-0.53}^{+0.74}$$_{-0.53}^{+0.51} \pm 0.07$  & $0.86 \pm 0.09 _{-0.15}^{+0.14} \pm 0.00\times10^{-2}$ &33.6/36\\
$10'<r<13'$ & $2.21_{-0.36}^{+0.67}$$_{-0.58}^{+0.77}$$_{-0.08}^{+0.11}$  & $0.40 \pm 0.06 _{-0.15}^{+0.12} \pm 0.00 \times10^{-2}$ & 45.5/49\\
$13'<r<16'$ & $2.13_{-0.55}^{+1.44}$$_{-0.60}^{+0.97}$$_{-0.10}^{+0.09}$  & $0.18 \pm 0.05 \pm  0.10 _{-0.01}^{+0.00} \times10^{-2}$ & N/A$^3$\\
\hline
\endhead
\multicolumn{4}{l}{\footnotesize $1$: $kT$ errors are shown in 90\% confidence level, with an order of statistical, CXB fluctuation and NXB fluctuation origins.}\\
\multicolumn{4}{l}{\footnotesize $2$: normalization in {\it apec} model, scaled to $\pi \times 20^2$~arcmin$^2$ flat region.}\\
\multicolumn{4}{l}{\footnotesize $3$: Fit to the outermost region is combined one to the NW and north. In addition, the foreground components were set free. See text for detail.}\\
\end{longtable}
\normalsize
\def\arraystretch{1}

\small
\def\arraystretch{1.25}
\begin{longtable}{llll }
\caption{Temperature and normalization of the spectra from the selected 3 regions fitted with the 2kT model.}
\label{tab:spectra2}
\hline
\hline
Region 	 & $kT_{\rm coool}$ and $kT_{\rm hot}$ keV & $norm_{\rm cool}$ and $norm_{\rm hot}$  & $\chi^2$/dof \\
\hline
Center  ~ $r<2'$ & $0.98_{-0.74}^{+0.34} \pm 0.00 \pm 0.00$  & $0.36_{-0.19}^{+0.17} \pm 0.01 \pm 0.00 \times10^{-2}$ & 215.0/177 \\
	& $3.99_{-0.31}^{+0.34}$$_{-0.11}^{+0.10} \pm 0.02$  & $5.84 \pm 0.29 \pm 0.13 \pm 0.00 \times10^{-2}$ &  \\
\hline
\hline
NW ~~~~~ $7'<r<10'$ & $0.76_{-0.59}^{+0.22}$$_{-0.02}^{+0.01} \pm 0.00$ & $0.08 \pm 0.03 \pm 0.00 \pm 0.00 \times10^{-2}$ & 89.5/98  \\
	& $2.81_{-0.33}^{+0.49}$$_{-0.38}^{+0.46} \pm 0.07$  & $0.75 \pm 0.07 \pm 0.10 \pm 0.00 \times10^{-2}$ &  \\
\hline
\hline
North ~~ $4'<r<7'$ &$0.99_{-0.76}^{+1.24}$$_{-0.02}^{+0.00} \pm 0.00$  & $0.17_{-0.11}^{+0.64}$$_{-0.02}^{+0.01} \pm 0.00 \times10^{-2}$ & 52.4/46 \\
	& $4.70_{-1.19}^{+2.19}$$_{-0.79}^{+0.59} \pm 0.10$  & $1.24_{-0.76}^{+0.30}$$_{-0.14}^{+0.16} \pm 0.00 \times10^{-2}$ &  \\
\hline
\endhead
\end{longtable}
\normalsize
\def\arraystretch{1}

\section{Temperature profiles}
\label{chap:kt_profile}

\subsection{Overall structure}

\begin{figure}
 \begin{center}
  \includegraphics[width=8cm]{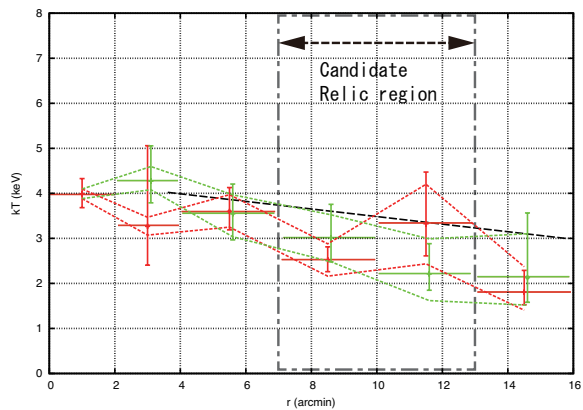}
 \end{center}
 \caption{Temperature profile of the ICM towards the NW (red) and the north (green) relic candidates. Error bars shown in cross are statistical 90\% confidence limit, while the thin dashed lines stand for the quadrature sum of both the CXB and NXB fluctuations. Note that plot of north regions (green) are artificially shifted by $+0'.1$ for clarity. Gray dot-dashed line from $7'$--$13'$ stands for the location of the candidate relics. Thick black dashed lines is the typical temperature profile given in Pratt et al. (2007) based on the XMM data, plotted by assuming an average temperature of $kT = 3.6$~keV.}
 \label{fig:kT_profile}
\end{figure}

The temperature profile as a function of the distance from the cluster center is shown in figure \ref{fig:kT_profile}. 
Average cluster temperature is $\sim 3.6$~keV, in good agreement with the XMM result (Solovyeva et al. 2008). 
Even though the surface brightness of the source is low,
thanks to the Suzaku low background, the ICM temperature was determined out to $r<16'$
with much better accuracy compared to XMM.
For example, in the region $7'<r<10'$ around the relic candidates, 
we obtained 90\% confidence statistical error of $\pm 0.27$~keV and $\pm 0.63$~keV for the NW and north regions, respectively,
while the XMM results in $7'<r<9'.5$ of north and NW co-added spectra
has an error of $\pm 1.8$~keV (converted into the 90\% confidence, from figure 8 of Solovyeva et al. 2008). 
Note that all errors in their paper is shown in 68\% confidence, while in this paper it is 90\%. 

The ICM temperature is in first approximation flat, with some symptom of fluctuation and marginal tendency for
getting lower to the outer radius, which is seen in many clusters (e.g. Pratt et al. 2007).
The $\sim 6$~keV hot region to the north at $4'<r<6'$ reported by Solovyeva et al. (2008) was not detected in Suzaku spectra,
although we do see milder jump, as discussed in the next subsection.
We quickly checked the XMM data and found there is a local
apparently hot region around source 2, which is almost excluded in our Suzaku analysis. Since the NXB of the XMM data
is already a bit high in this region, we did not go into farther detail on the XMM data and focus on our Suzaku data in this paper.

\subsection{Search for shock symptoms around the NW relic candidate}
\label{sec:search_shock}

Here we focus on the temperature structure around the two relic candidates.
Since majority of the north relic candidate region is masked out by source 3, here we focus on the NW arc.

Although there is no evidence for $\sim 6$~keV hot region at around $4' < r< 6'$, there exists a temperature 
jump at $r\sim7'$, from 3.60~keV at $4'<r<7'$ to 2.52~keV at $7'<r<10'$ annulus (see figure \ref{fig:kT_profile}), 
which is qualitatively consistent with the XMM result.
Temperature ratio of the outer region compared to the inner one is calculated to be $1.43\pm0.24$. 
Here the error is at statistical 90\% confidence level, i.e. 1.65$\sigma$. If we include the CXB fluctuation effect, the error rises to 0.35.
When corrected for the ``average ICM temperature gradient'' (e.g. Pratt et al. 2007), it become $1.36\pm 0.35$.
Thus, the temperature rise is significant right at $1.65\sigma$ (or at one-side 5\% confidence level) judging from the 1kT fit.

Looking at figure \ref{fig:kT_profile}, however, the temperature profile (shown in red) can be interpreted as a ``dip'' at $7'<r<10'$ annulus.
Thus, it is natural to conclude there is some ``cooler'' gas at this region, rather than 
assuming hotter gas in the inner.
The 1kT fit to the NW $7'<r<10'$ spectra actually shows small positive residual at around 0.8--1.0~keV and negative around 1.5~keV.
With the 2kT model, as shown in table \ref{tab:spectra2}, 
we have $kT_{\rm cool} = 0.76^{+0.22}_{-0.59}$~keV and $kT_{\rm hot} = 2.81^{+0.68}_{-0.51}$~keV, respectively 
(all errors are shown in quadrature sum).
The $\chi^2/$dof improved to 89.5/98 from 105.0/100 of the 1kT fit, and {\it f}-test shows NHP of 0.06\%.
Thus, the spectra can be well explained by a combination of minor ($\sim10$\% in its normalization) 
$kT_{\rm cool}\sim 0.8$~keV and major $kT_{\rm hot} \sim 2.8$~keV components.
In other words, there is no temperature jump if we think the hot component is the main ICM.

Spectra from other regions including relic candidates (both NW and north, at annuli of $7'<r<10'$ and $10'<r<13'$) shows similar residual
but with less significance. Improvement of fit with the 2kT model in view of {\it f}-statistics is only about 1--7\% in NHP, 
and what is more the 1kT fit itself is acceptable in 90\% confidence level.
So the cool component will be existing all around the candidate NW relic region, but with only marginal evidence with the Suzaku data.

The $kT \sim 0.8$~keV cool component is also suggested in the north $4'<r<7'$ spectra. As already mentioned, 
the 1kT model fit to the data gave a marginal NHP of 9.9\% and the residual spectra has a soft excess.
With the 2kT model, the fitting improved as shown in table \ref{tab:spectra2}, and {\it f}-test shows significant NHP of 0.04\%.
The normalization of the cool component here is $\sim 14$\% of the hotter one.
Thus, the minor cool component is also suggested to be mixed in the ICM at the region between the north relic candidate and the cluster center.

Another temperature jump candidate is at the outer rim of the NW relic candidate at $r\sim 13'$,
from 3.3~keV at $10'<r<13'$ to 1.8~keV at $13'<r<16'$ annulus. With only statistical error,
the ratio is $1.83\pm 0.53$, and with the CXB fluctuation it becomes $1.83\pm 0.94$. 
Again corrected for the ``ICM temperature gradient'', it becomes $1.71\pm 0.94$.
This is $\sim 1.25 \sigma$, meaning that the possibility the temperature is ``higher'' in the inner annulus is 89\%.
This result marginally prefers the temperature jump at the outer rim of the NW relic candidate, but not significant enough to conclude on it.
Note that we see no symptom of similar temperature jump at the north regions.

\subsection{Entropy profile}

\begin{figure}
 \begin{center}
  \includegraphics[width=8cm]{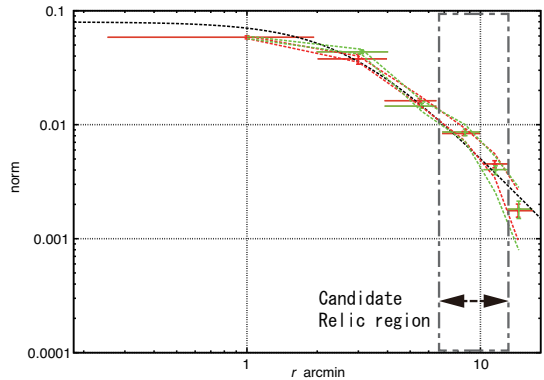}
 \end{center}
 \caption{Normalization of the spectral fitting towards the NW (red) and the north (green) relic candidates. Error bars are in the same format as in figure \ref{fig:kT_profile}. Black dash line is the scaled  $\beta$ model. See text for detail. }
 \label{fig:S_profile}
\end{figure}

\begin{figure}
 \begin{center}
  \includegraphics[width=8cm]{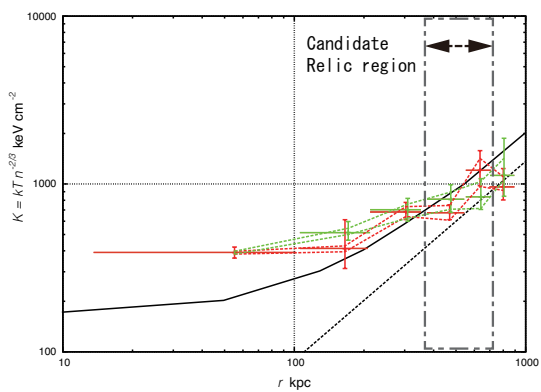}
 \end{center}
 \caption{Entropy profile estimated by assuming the $\beta$ model distribution of the ICM. Profile to the NW (red) and north (green) are shown. Error bars are in the same format as in figure \ref{fig:kT_profile}. For comparison, average profile of the 109 high central entropy ($K_0 > 50$~keV cm$^2$) clusters by Cavagnolo et al. (2009) is  shown in solid black line. The profile of so called gravitational accretion heating model from Voit et al. (2002) is also shown (dashed black).  }
 \label{fig:K_profile}
\end{figure}

With the ICM temperature and density obtained, we then calculated the astrophysical entropy profile,
given as $K = \frac{kT}{(n_e)^{2/3}}$ (e.g. Ponman et al. 1999). Here, $n_e$ is the electron density.
Since our Suzaku observation only covers the northern portion of the cluster, 
here we assume the $\beta$ model gas distribution with a core radius $r_c = 2'.82$ and $\beta = 0.52$, 
provided by Neumann \& Arnaud (1999) obtained from the ROSAT PSPC analysis.
As shown in figure \ref{fig:S_profile}, the fitted {\it apec} normalization profile matches well with 
the $\beta$ model with residuals less than $\sim 30$\%.
By scaling the $\beta$-model to the {\it apec} normalization profile, 
the central $n_e$ was derived as $n_e (r=0) = 1.25\times10^{-3}$~cm$^{-3}$.

The obtained entropy profile is shown in figure \ref{fig:K_profile}. Error bars include both of them from the $n_e$ and $kT$ estimations.
Central entropy of $K_0 = 400$~keV cm$^{-2}$ at around 50~kpc from the cluster center
is one of the highest among the clusters with this temperature.
Actually, there is only one object with entropy higher than this in the $T_{\rm cluster} < 4$~keV panel of
the entropy profiles (figure 5) of Cavagnolo et al. (2009), compiled from the Chandra data of 239 clusters with various temperature.
Because the center spectra are fitted with 2kT model and we only employed the hotter one as the ICM component,
the central entropy would be a little overestimated. However, when applying the 1kT fit results and perform the same calculation,
we get a central entropy of $K_0 = 320$~keV cm$^{-2}$, which is still high.

In the profile, apparent ``dip'' in the NW direction (red lines) at around $r\sim 450$~ kpc and candidate jump at $\sim 800$~kpc
both reflects the temperature structure discussed in the last subsection.
Over all, the entropy is high and flat, with no significant structure. 
This is consistent with the X-ray image being relatively circular, as well as its general lack of strong temperature structure.

\subsection{Upper-limits on the inverse Compton emission}
\label{sec:IC}

The diffuse radio sources are presumably synchrotron emission by GeV electrons interacting with 
$\sim \mu$G magnetic field in the ICM. The same electrons scatter the Cosmic microwave background up to
the X-ray energy band, i.e. so called inverse-Compton (IC) emission.
Since the X-ray spectra around the radio sources are well modeled with thermal emission,
here we estimate the upper limit on the emission. 

\begin{figure}
 \begin{center}
  \includegraphics[width=8cm]{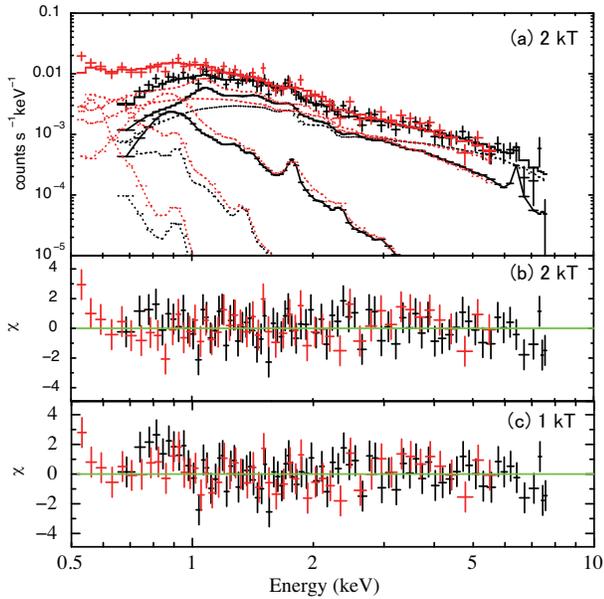}
 \end{center}
 \caption{(a) Spectra obtained around the NW relic candidate region, fitted with the 2kT model, and (b) its residual. Panel (c) is the residual with the 1kT model.}
 \label{fig:1kT+PL}
\end{figure}

Again, we focus on the NW relic and select a region with a radius of $3'.5$ around it. 
Region $2'$ around source 4 is also masked out.
When fitted with the 1kT model, we obtain an acceptable result with $kT\sim 2.63_{-0.26}^{+0.32}$$_{-0.36}^{+0.37} \pm 0.05$~keV 
and $\chi^2/$dof$ = 142.0/126$. However, as already suggested in the annulus spectra, the residual around 0.8--1~keV exists,
and 2kT model gives significantly better fit with $kT_{\rm cool} = 0.86 \pm 0.14 \pm 0.00 \pm 0.00$~keV, 
$kT_{\rm hot} = 3.14_{-0.45}^{+0.65}$$_{-0.42}^{+0.51}$$_{-0.09}^{+0.08}$~keV and $\chi^2/dof = 118.0/124$. 
Spectra of the 1kT and 2kT model fit is shown in figure \ref{fig:1kT+PL}.

The IC component will have a power-law like spectra. 
Here we assume its photon index to be $\Gamma = 2.0$ (fixed), i.e. flat in $\nu F \nu$ plot, for simplicity.
Although the value observed in 1.4~GHz radio (Feretti et al. 2006) is a bit softer,
they are still consistent within the error. Assuming 1~$\mu$G magnetic field, electrons scattering 8~keV
X-rays corresponds to those emitting 38~MHz radio, well below the observed 1.4~GHz band.
Thus, assuming a little harder spectral index there is natural.
Unfortunately, the hotter component of the 2kT fit has very similar shape to the $\Gamma = 2.0$ power-law.
Actually, if we replace it by a power-law with $\Gamma$ free, it was derived as $\Gamma = 2.1_{-0.2}^{+0.8}$ 
(errors are mostly from the CXB fluctuation).
Nonetheless, we fitted the spectra with a fixed $\Gamma$ ($= 2.0$) power-law in addition to the 2kT model
and estimated the upper-limit flux of the former component as $0.8\times 10^{-13}$~erg s$^{-1}$ cm$^{-2}$ at 2--10 keV.

By integrating the power-law energy distribution of electrons in Lorentz factor of $500<\gamma<4\times10^4$,
and assuming the relic has a spherical shape with a radius of 190~kpc ($3'.5$),
the electron energy density becomes $U_e < 0.2$~eV cm$^{-3}$.
This is not well constrained, compared to the  thermal energy density of 
$\sim 1.1$~eV cm$^{-3}$ (calculated assuming $n_e \sim 2\times10^{-4}$ cm$^{-3}$ and electron-ion number ratio of 1.2). 
Combining the 1.4~GHz radio flux ($61\pm5$~Jy, Feretti et al. 2006) and the hard X-ray flux,
we obtain the lower limit magnetic field strength of $>0.5$~$\mu$G, which is consistent with 
the equipartition field of 0.9~$\mu$G (Solovyeva et al. 2008). 

\section{Discussion}

\subsection{Short summary of Suzaku results}

We analyzed the Suzaku deep (84.4~ks) observation data of Abell 548W and
measured the ICM properties out to $r=16'$ (or 880~kpc) from its center, well beyond the two relic candidates.
When estimated from the ``average'' ICM temperature of 3.6~keV, $r_{200}$ and $r_{500}$ become 1.3~Mpc and 860~kpc, respectively.
Here we used the data provided in Arnaud et al. (2005) and estimate
$r_{200} = 704\times \sqrt{kT/{\rm keV}}$~kpc and $r_{500} = 452\times \sqrt{kT/{\rm keV}}$~kpc.
Thus, our observation range reaches 2/3 of $r_{200}$ and slightly exceeds $r_{500}$ derived under hydrostatic assumption.
Out to this radius, the ICM morphology, temperature and entropy do not show strong structure,
with marginal evidence for small temperature variation.

The ICM temperature is $\sim 4$~keV at its center, and $\sim 2$~keV at the outermost regions along the two relic candidates.
We also observe a temperature ``dip'' around the NW relic candidate,
which is understood as a $\sim 1$~keV cold gas mixed with the $\sim 3$~keV ICM emission.
Outer rim of the NW relic shows marginally higher temperature than those of the ICM outside,
consistent with the relic candidate being located at the shock edge.
Its significance is marginal, i.e. only 89\% confidence, not strong enough to conclude its existence.
Astrophysical entropy calculated from the ICM density and temperature
reaches 400~keV~cm$^{-2}$ at around 50~kpc from the cluster center, which is among the highest of clusters.
This value directly reflects the low surface brightness nature of the cluster.

\subsection{ICM properties in view of gas mass fraction}

\begin{figure}
 \begin{center}
  \includegraphics[width=8cm]{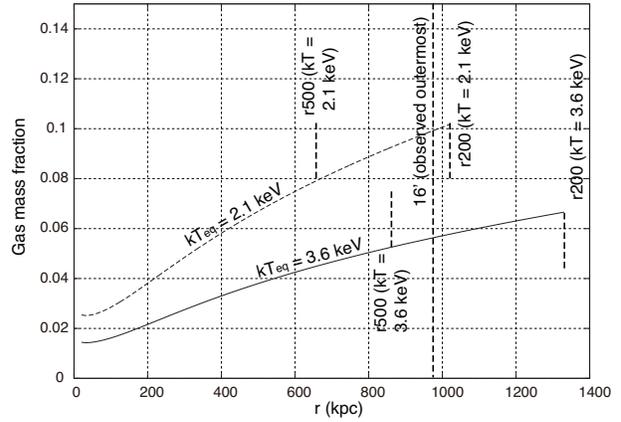}
 \end{center}
 \caption{Ratio of gas mass to total mass calculated out $r_{200}$, estimated by using the ICM temperature. Solid line indicates that derived from the total mass assuming hydrostatic equilibrium with $kT = 3.6$~keV. Dotted line indicates those obtained by assuming the hydrostatic equilibrium temperature to be 2.1~keV, i.e. heated by a factor of 1.7. For reference, $r_{500}$ is also shown in each plot. Vertical dot dashed line stands for $r=880$~kpc, which corresponds to the outer bounds $r = 16'$ of our data analysis.}
 \label{fig:fbaryon}
\end{figure}

Because the X-ray image is relatively circular, we here assume that the cluster itself is spherically symmetric for simplicity.
The ICM distribution is well modeled with the $\beta$ model, and gas mass integrated within a radius $r$ can be estimated.
Assuming hydrostatic equilibrium, we can also calculate the total mass of the cluster.
We then derived the ``gas-mass fraction'' ($f_{\rm gas}$) using the ratio of these two, calculated out to $r_{200}$.
As shown in figure \ref{fig:fbaryon}, $f_{\rm gas}$ is derived as $6.7$\% at 1.3~Mpc ($\sim r_{200}$ for $kT = 3.6$~keV).
Gas mass integrated out to $r_{200}$ is derived as $M_{\rm gas}(r_{200}) \sim 1.9\times10^{13} M_{\odot}$, while
the hydro-statically estimated total mass is $M_{\rm tot}(r_{200}) \sim 29\times10^{13} M_{\odot}$. 
The derived  $f_{\rm gas}$  is less than a half of the value generally reported in other clusters.
For example, Walker et al. (2013) derived $f_{\rm gas} \sim 0.15$ at $r_{200}$ in Centaurus cluster, which has a temperature of $\sim 3$~keV, 
similar to that of Abell 548W. Allen et al. (2008) also showed $f_{\rm gas} \sim 0.15$ in the analysis of 42 clusters with temperatures higher than 5~keV.

To compare $f_{\rm gas}$ with many other clusters, we also derived the value at $r_{500}$.
Then we have $M_{\rm gas}(r_{500}) \sim 0.98\times10^{13} M_{\odot}$ and $M_{\rm tot}(r_{500}) \sim 19\times10^{13} M_{\odot}$, resulting in 
$f_{\rm gas} = 0.052$. According to Pratt et al. (2007), who analyzed 31 nearby clusters, 
the averaged gas mass fraction $\bar{f}_{\rm gas}(r_{500})$ is $0.09$
for a cluster with $M_{\rm tot}(r_{500}) = 19\times10^{13}~M_{\odot}$ (see figure 8 of the paper).
Thus, also at $r_{500}$, Abell 548W shows slightly more than a half of $f_{\rm gas}$ of that of the ordinary cluster.
Note that there are a few clusters at around $M_{\rm tot}(r_{500}) \sim 19\times10^{13}~M_{\odot}$ which showed $f_{\rm gas}(r_{500})$
as small as 0.05 in their plot. Because these plots actually include the three LSB clusters presented at the beginning (Abell 2399, Abell 3771 and Abell 2328),
it means that Abell 548W has a typical $f_{\rm gas}(r_{500})$ of LSB clusters. 
While the morphologies of the three LSB clusters are disturbed, that of Abell 548W is rather circular, 
which makes this cluster peculiar.

In this scenario we assumed that the ICM of this cluster is in hydrostatic equilibrium and simply its $f_{\rm gas}$ is small.
As already noted in, e.g. Ota et al. (2014), the ICM cannot be radiatively cooled down to the 
``ordinary''  entropy within $10^{10}$~years, and thus this cluster remains LSB 
for a long period.

The fact that the XMM image has $\sim 2$ X-ray peaks possibly associated with the 2 elliptical galaxies in its center,
suggests that the cluster is dynamically young.
In addition, as already noted, its galaxy velocity dispersion $\sigma_V =1300$~km s$^{-1}$
is too high as a $kT\sim 3.6$~keV cluster. Based on the redshift distribution of 193 galaxies
in the Abell 548W region, Solovyeva et al. (2008) interpreted it as a 
mixture of 2 clusters each with $\sigma_V = 700$ and $900$ ~km s$^{-1}$,
merging with a relative velocity of $\sim 1500$~km s$^{-1}$.

We then consider a scenario that the ICM temperature is heated up by cluster merger.
In this case, hydrostatic equilibrium is not taking place, and the relic candidates
can be interpreted as the shock front propagating outward.
Let's here assume that the ICM is heated up by a factor of 1.7, i.e. it will settle down  to 2.1~keV {\it after} final relaxation (dynamically, not by cooling). In this case, total mass derived in the last paragraph is overestimated by the same factor. As shown in figure \ref{fig:fbaryon},  {\it real} $r_{500}$ becomes as small as 650~kpc with $M_{\rm tot}(r_{500}) \sim 7.9\times 10^{13}~M_{\odot}$. Then, $f_{\rm gas}(r_{500})$ reaches 0.08, which is the typical value shown in figure 8 of Pratt et al. (2007). 
In the merger scenario, the cool component seen in a few regions can be understood as remnants of the pre-shock low entropy gas.

\subsection{Line-of-sight major merger scenario}

The simplest toy-model is a line-of-sight, 1:1 major merger, with the pre-shock temperature of $\sim 1.8$~keV. 
Pre-shock sound velocity becomes $v_s \sim 680$~km s$^{-1}$ and with a colliding velocity of 1500~km s$^{-1}$, 
it can generate a shock with Mach $\sim 2$, and hence post-shock temperature of 3.6~keV from Rankine-Hugoniot conditions.
Specifically, defined {\it inward} to the center of gravity of the system,
pre-shock bulk velocity is assumed to be $750$ ($=1500/2)$~km~s$^{-1}$, that of post-shock 0~km~s$^{-1}$, 
and shock plane velocity {\it outward} $600$~km~s$^{-1}$.
With these parameters, the Mach number becomes $(750 + 600~{\rm km~s^{-1}})/680~{\rm km~s^{-1}} = 2.0 $ and thus
the temperature ratio 2.1.

Here we consider that this cluster (or two groups) is (are) right at the middle of initial heating phase (see, e.g. figure 5 of Ricker and Sarazin 2001).
In the following adiabatic expansion, the X-ray luminosity will follow
$L_X \propto n_e n_i T^{0.5} V = (n_i V) n_e T^{0.5} \propto M_{\rm gas} T^2$.
Here, $n_i$ is ion number density, $M_{\rm gas}$ is total gas mass, and entropy conservation of $K = T/(n_e)^{2/3}$ is applied.
With the cluster total mass to be doubled after merger, its future relaxation temperature will be $\sim$ 2.8~keV,
assuming the $M$--$T$ relation ($M \propto T^{1.6}$ by Vikhlinin et al. 2006).
As the merger moves to later stages, the decreasing temperature will cause the luminosity to get dimmer as $\propto T^2$. 
Since $L_X(r_{500}) \propto T(r_{500})^{2.7}$ in Pratt et al. (2007), luminosity deficit will be slightly relaxed as the cluster settles down.
In addition, larger scatter in $L_X$ at lower temperature make the peculiarity of this object further relaxed.
In other words, this cluster in future will look like {\it one of many low $L_X$ groups of galaxies} 
sometimes seen in the $kT$--$L_X$ plot.

Although the X-ray properties of Abell 548W could be understood if it is a major merger of (relatively large) galaxy groups,
the merging velocity of $1500$~km s$^{-1}$ itself is rather high, and we need to consider its origin in our future work.
What is more, general lack of strong inhomogeneity in both the X-ray morphology and temperature structure 
requires a finely tuned merger model, e.g. merger axis perfectly aligned to the line of sight, and so on.
Other exotic possibilities, such as overheating by AGN feedback and inherent baryon fraction deficit, 
still cannot be ruled out with current observational results.

\section{Summary}

Suzaku deep (84.4~ks) observation of Abell 548W detected
the ICM emission out to $r=16'$ (or 880~kpc) from its center, well beyond the two relic candidates,
and measured the ICM temperature for the first time out to this radius.
The ICM morphology, temperature and entropy do not show strong structure,
while marginal evidence for small temperature variation is observed.
The hot ($\sim 6$~keV) component detected with XMM (Solovyeva et al. 2008) was
not confirmed, although the contaminating point source (source 1) makes it difficult for Suzaku 
to clearly distinguish the inconsistency.
Central entropy of the ICM is among the highest in a cluster with this temperature, as well.

At the NW candidate relic region, symptom of relatively cool ($\sim 1$~keV) component mixed  with
the $\sim 3$~keV ICM emission is detected.
In addition, marginal temperature jump at the NW relic rim is suggested.
If this is the case, the radio sources are consistent with being relics activated with merger shock.

When assuming hydro-static equilibrium, the gas-mass fraction ($f_{\rm gas}$) of the cluster
is estimated to be 0.067 at $r_{200}$ and 0.052 at $r_{500}$, which both are about a 
half of the value generally seen.
Considering these observational properties, 
a merging cluster scenario of two relatively large ($kT\sim 1.8$~keV) galaxy groups is discussed. 
Although these parameters can explain the high entropy nature of the cluster,
finely tuned model to address both the high entropy and featureless and apparently circular X-ray properties at the same time will be needed.
In other words, if this cluster is a major merger, the merging axis shall be almost completely parallel to the line-of-sight.

\section*{Acknowledgement}

KN, MT and MF are supported in part by JSPS KAKENHI Grant Number 15H03639, 26400218 and 15K05080, respectively.

\end{document}